\documentclass[11pt]{article}

\usepackage[utf8]{inputenc}

\usepackage{caption}
\usepackage{subcaption}

\usepackage{geometry} 
\usepackage{hyperref}
\usepackage{amsmath}
\usepackage{graphicx}

\usepackage[square, numbers]{natbib}
\usepackage{booktabs}
\usepackage{array}
\usepackage{paralist}
\usepackage{verbatim}
\usepackage{subcaption}

\usepackage{adjustbox}

\usepackage{fancyhdr}
\usepackage{amsmath}

\usepackage{sectsty}
\allsectionsfont{\sffamily\mdseries\upshape}

\newcommand{\be}{\begin{equation}}
\newcommand{\ee}{\end{equation}}

\usepackage[table]{xcolor}
\usepackage{etoolbox}
\usepackage{pgf}

\definecolor{high}{HTML}{FF3333}
\definecolor{low}{HTML}{ECFF33}
\newcommand*{\opacity}{90}

\newcommand*{\minval}{1.0}
\newcommand*{\maxval}{1.23}

\newcommand{\gradient}[1]{
    \ifdimcomp{#1pt}{>}{\maxval pt}{#1}{
        \ifdimcomp{#1pt}{<}{\minval pt}{#1}{
            \pgfmathparse{int(round(100*(#1/(\maxval-\minval))-(\minval*(100/(\maxval-\minval)))))}
            \xdef\tempa{\pgfmathresult}
            \cellcolor{high!\tempa!low!\opacity} #1
    }}
}
\begin{document}

\title{Portfolio Optimization Rules beyond the Mean-Variance Approach}

\author{Maxime Markov\thanks{Corresponding Author: \href{markov@theory.polytechnique.fr}{markov@theory.polytechnique.fr}}  \,\, and Vladimir Markov}

\date{\today}

\maketitle

\begin{abstract}
In this paper, we revisit the relationship between investors' utility functions and portfolio allocation rules. We derive portfolio allocation rules for asymmetric Laplace distributed $ALD(\mu,\sigma,\kappa)$ returns and compare them with the mean-variance approach, which is based on Gaussian returns. We reveal that in the limit of small $\frac{\mu}{\sigma}$, the Markowitz contribution is accompanied by a skewness term. We also obtain the allocation rules when the expected return is a random normal variable in an average and worst-case scenarios, which allows us to take into account uncertainty of the predicted returns. An optimal worst-case scenario solution smoothly approximates between equal weights and minimum variance portfolio, presenting an attractive convex alternative to the risk parity portfolio. We address the issue of handling singular covariance matrices by imposing conditional independence structure on the precision matrix directly.  Finally, utilizing a microscopic portfolio model with random drift and analytical expression for the expected utility function with log-normal distributed cross-sectional returns, we demonstrate the influence of model parameters on portfolio construction. This comprehensive approach enhances allocation weight stability, mitigates instabilities associated with the mean-variance approach, and can prove valuable for both short-term traders and long-term investors.
\end{abstract}

\section{Introduction}

Investors and portfolio managers constantly strive to optimize their investment strategies in order to maximize returns while minimizing risk. This objective is formalized in the portfolio optimization rules, which facilitate the selection of a set of stocks that yield the highest returns possible considering the risk involved.

Since the introduction of the mean-variance approach in 1952~\cite{Markowitz:1952}, the field has rapidly evolved to include different risk and expected return models into the framework. Despite this, simplistic rules, such as equal weight or equal volatility, have demonstrated resilience in practice \cite{Demiguel:2009}, while many advanced formulas remain underutilized. The framework's idealistic presumptions are the issue. 
The mean-variance approach is based on the assumption that stock returns are Gaussian, whereas real-world stock returns often exhibit fat tails and skewness toward negative returns. Although expected returns display some predictability over the long term because of their correlation with fundamental data, mid- and short-term returns are inherently unpredictable. Practically, the framework's reliance on inverting the low-rank covariance matrix introduces numerical instabilities to an already extensive list of assumptions that are often invalid in a noisy and non-stationary world. Critiques of modern portfolio theory span from popular authors~\cite{Taleb:2008} to academics~\cite{Cochrane:2021}.

In this paper, we propose a set of models to enhance the quality of asset allocation rules by incorporating uncertainty of the expected returns in average and worst-case scenario and improving stability of covariance matrix estimation by imposing block constraints on a precision matrix. First, we examine historical returns of major stock indexes on daily, weekly, and monthly time scales, demonstrating that the asymmetric Laplace distribution (ALD) provides a reasonable fit. We derive empirical location ($\mu$), scale ($\sigma$), and skewness ($\kappa$) parameters as well as their scaling laws. Second, we derive allocation rules for ALD returns and systematically compare them to the mean-variance approach. We demonstrate that in the limit of small $\frac{\mu}{\sigma}$, the Markowitz contribution is accompanied by a non-negligible skewness term. Furthermore, the fat tails impose a cut-off on risk by saturating allocation weights for stocks with high expected returns.  Third, we study the proposed models when the expected return $\mu$ is a Gaussian random variable $\mu \sim N(\mu_0,\sigma_0)$, with the main objective to find allocation rules when uncertainty $\sigma_0$ is much larger than $\mu_0$. We derive allocation rules in an average scenario by marginalization over $\mu$, and in the worst-case scenario. We demonstrate that in the limit of noisy prediction of the expected return $\frac{\mu_0}{\sigma_0}\to 0$, the allocation rule of the average scenario resembles the mean-variance solution with regularization, where $\sigma_0$ acts as a regularizer (shrinkage). In the worst-case scenario, the optimal solution smoothly approximates between equal weights and a minimum variance portfolio, thus providing an attractive \mbox{\it{convex}} alternative to a risk parity (RP) portfolio. Fourth, we investigate the integration of the graphical structure dictated by conditional independence of investments into portfolio allocation rules. We demonstrate that this structure can be naturally incorporated into the estimation of the precision matrix. This approach mitigates the sampling error in large covariance matrices, reduces the number of nonzero elements, and enables a more stable inversion of covariance or precision matrices. Finally, we derive allocation rules for log-normal distributed returns, which serve as a base model for long-term cross-sectional returns~\cite{Markov:2023}.

We anticipate that the proposed framework and methods will prove beneficial for practitioners in the realm of portfolio construction, potentially improving or superseding the mean-variance and risk parity approaches. 

\section{Risk perception, utility theory, and portfolio allocation rules}

Portfolio optimization task can be defined in many ways, as it is contingent upon the diversity of risk perception and the distribution of investment outcomes. In the standard economic approach, asset allocations are derived from maximizing the expected value of the investor's utility function. The most commonly used utility models in finance are the Constant Absolute Risk-Aversion (CARA) and Constant Relative Risk-Aversion (CRRA) models~\cite{Campbell:2001}.
The utility functions $U_a$ for CARA and $U_r$ for CRRA with respect to the investment outcome (return) $x$ and with $a$  and $\gamma$ being the risk aversion parameters can be expressed as follows:
\be
 U_a(x,a) = 
    \begin{cases}
       \frac{1-\exp(-a x)}{a}, &  \mbox{if }a \neq 0 \\
        x, &  \mbox{if }  a = 0
    \end{cases}, \,\, \,
 U_r(x,\gamma) = 
    \begin{cases}
        \frac{x^{1-\gamma}-1}{1-\gamma}, &  \mbox{if } \gamma > 0, \gamma \neq 1 \\
        \ln{x} & \mbox{if } \gamma = 1
    \end{cases}
\label{eq:utility}
\ee
To determine the optimal portfolio weights, we maximize the expected value of the utility function $U(x,a)$ with respect to $w$:
\be
w=arg \max_w E_P[U(x,a)]=arg \max_w \int dx\,\, U(x,a) P(x)
\label{eq:potfolio-opt}
\ee
where $P(x)$ is a distribution of outcome $x$.  Mean variance approach assumes that returns $P(x)$ are Gaussian. In Section 3, we show that ALD provides a better fit while capturing fatter tails and skewness of empirical returns.

\section{Empirical distribution of stock returns on daily, weekly, and monthly time scales}
\label{sec:empirical}
In this section, we examine the empirical return distribution $P(x)$ of 38 stock index time series over a period from January 1, 2002, to December 31, 2022, on daily ($D$), weekly ($W$), and monthly ($M$) time scales.  We group indexes according to their geographical location into several groups, including the United States, Europe, Asia-Pacific (APAC), Japan, and BRIC (Brazil, Russia, India, and China) countries. We also study 11 sectors of S\&P 500 GICS Level 1 indexes to better understand the role of heterogeneity within the S\&P 500 index. We assume that index returns serve as a proxy for individual stock returns over the chosen time frame and market.

All time series exhibit non-normal characteristics, specifically fat-tails and skewness in return distribution. To account for these properties, we use the ALD with the probability density function (PDF) given by~\cite{Kotz:2001}:
\be
p(x,\mu,\sigma,\kappa)=\frac{\sqrt{2}}{\sigma}\frac{\kappa}{1+\kappa^2} \exp{\left(-\sqrt{2} \frac{(x-\mu)}{\sigma}  s \kappa^s \right)}
\label{ald}
\ee
where $s=sgn(x-\theta)$,  $\mu$ is the location parameter, $\sigma$ is the scale parameter and $\kappa$ is the skewness parameter. The case of zero skewness corresponds to $\kappa=1$. The skewness toward negative returns corresponds to $\kappa>1$. Note that there is a widely used \textit{classical} ALD, which differs from the ALD in Eq.~\ref{ald} by $\sqrt 2$ in the definition of scale $\sigma$.

\begin{figure}[t]
    \centering
     \begin{subfigure}[b]{0.49\textwidth}
         \centering
         \includegraphics[width=1.1\textwidth]{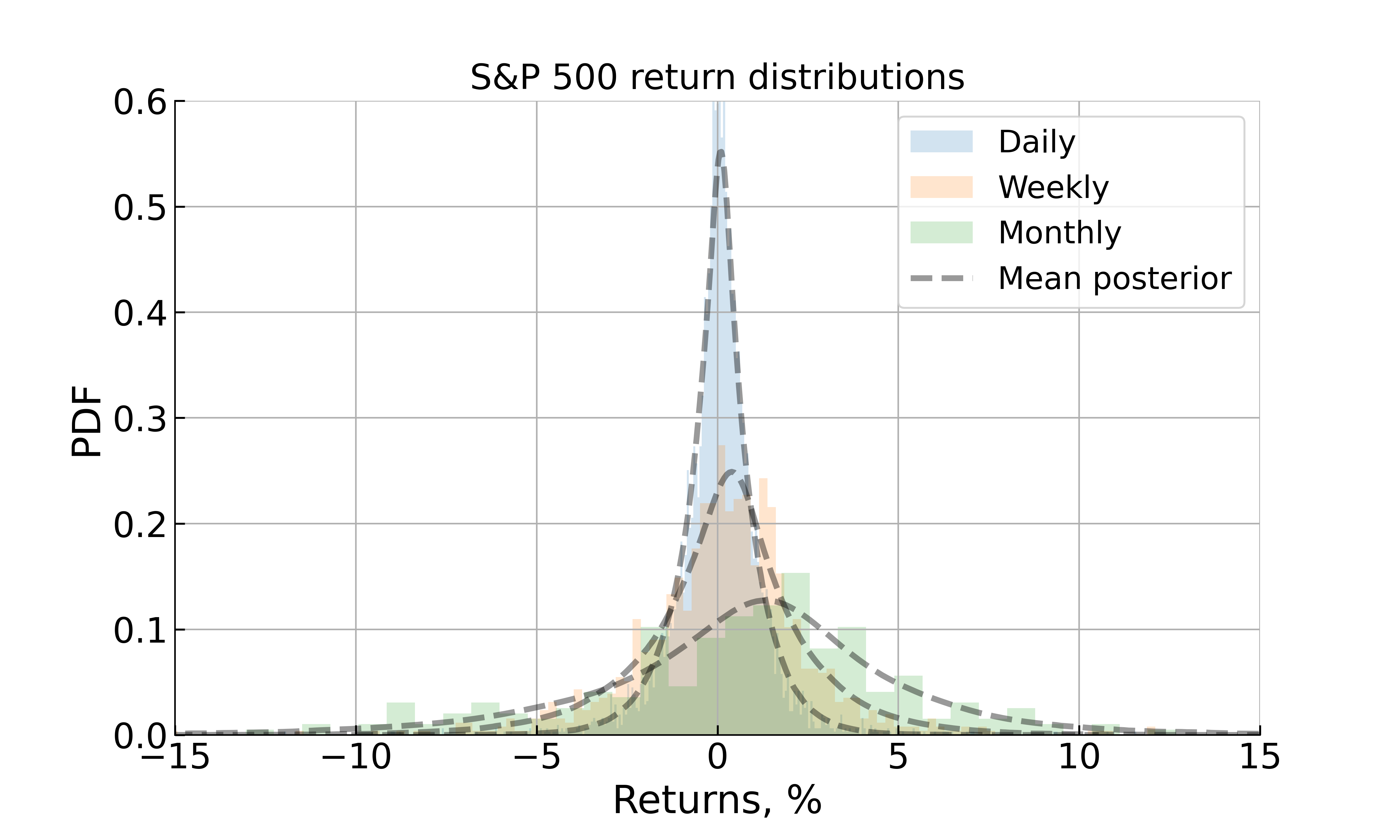}
     \end{subfigure}
     \hfill
    \begin{subfigure}[b]{0.49\textwidth}
         \centering
         \includegraphics[width=1.1\textwidth]{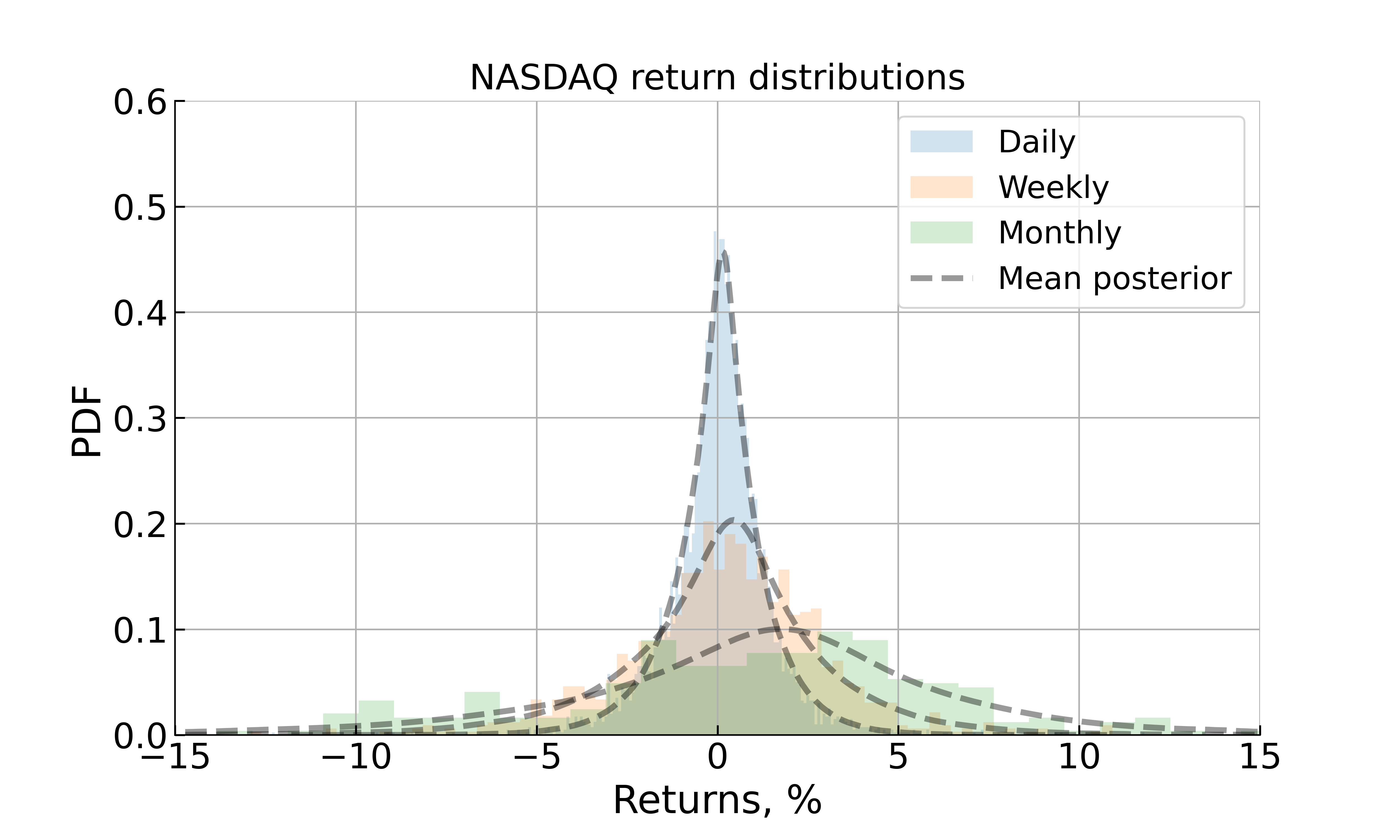}
     \end{subfigure}
    \caption{Probability density functions (PDF) of empirical daily (blue), weekly (orange), and monthly (green) returns of S\&P 500 and NASDAQ indexes. Dashed lines show the mean posterior of the ALD fitted with the MCMC method. Higher period distributions show greater asymmetry biased toward negative returns.}
    \label{fig:distributions}
\end{figure}

The empirical PDF of the daily (blue), weekly (orange), and monthly (green) returns of S\&P500 (SPX) and  NASDAQ (CCMP) indexes are shown in Figure~\ref{fig:distributions}. It can be noticed that the daily returns have a sharp peak at zero. As the time-frame increases, the distribution becomes broader and more skewed toward negative returns. In order to quantify this effect, we fit the historical return data with the ALD. The distribution parameters are estimated using Markov Chain Monte Carlo (MCMC) method as implemented in PyMC3~\cite{Salvatier:2016}. The dashed lines represent the mean posterior of the fits.

We repeat the same procedure for all indexes and their distribution parameters $\mu$, $\sigma$, and $\kappa$ for the daily, weekly, and monthly returns are summarized in Table~\ref{table:ALD-fit}. To highlight the effect of increasing skewness, we have colored the columns with $\kappa$ values for each return.

Given the dependence of the distribution parameters for the indexes presented in Table~\ref{table:ALD-fit}, we seek to extract the scaling laws for the averaged parameters.  To achieve this, we utilize a linear regression method on a log-log scale. By analyzing the slope of a linear curve, we obtain the following scaling laws:
\begin{equation}
\mu_T \sim T^{0.78},\,\, \sigma_T \sim T^{0.46}
\end{equation}
where $T$ represents the measured period in trading days. Both the location and scale parameters exhibit a positive slope. In contrast, the behavior of the skewness parameter $\kappa$ is idiosyncratic, and it may exhibit non-monotonic behavior. An increase in $\kappa$ signifies the persistence of negative returns across the investigated time scales. Therefore, it would be prudent to select $\kappa$ based on the time scale that corresponds to the anticipated portfolio holding period.

We observe a linear relation $\frac{\mu_T}{\sigma_T}=a*(\kappa_T-1)+b$, with  parameters $a_D = 2.8$,  $b_D = 0.1$, $a_W = 2.9$,  $b_W = 0.1$, and $a_M = 2.8$,  $b_M = 0.3$ for daily, weekly, and monthly data as summarized in Table 1. The regression chart is shown in Appendix D.   

\begin{table}[p]
\centering
\begin{adjustbox}{width=0.95\textwidth}
\begin{tabular}{|l|rrr|rrr|rrr|}
\hline
 & \multicolumn{3}{l|}{Daily} & \multicolumn{3}{l|}{Weekly} & \multicolumn{3}{l|}{Monthly} \\
 \hline
 &  $\mu$ &  $\sigma$ &  $\kappa$ &     $\mu$ &  $\sigma$ &  $\kappa$ &     $\mu$ &  $\sigma$ &  $\kappa$ \\
 \hline
\multicolumn{10}{l}{\textbf{US}} \\
\hline
SPX     &  0.10 &  0.57 &  \gradient{1.04} &  0.41 &  1.19 &  \gradient{1.08} &  1.48 &  2.17 &  \gradient{1.13} \\
CCMP    &  0.15 &  0.69 &  \gradient{1.06} &  0.52 &  1.46 &  \gradient{1.08} &  1.92 &  2.76 &  \gradient{1.12} \\
RIY     &  0.10 &  0.57 &  \gradient{1.04} &  0.43 &  1.20 &  \gradient{1.09} &  1.56 &  2.20 &  \gradient{1.13} \\
RTY     &  0.16 &  0.76 &  \gradient{1.06} &  0.46 &  1.59 &  \gradient{1.07} &  1.73 &  2.81 &  \gradient{1.11} \\
RAY     &  0.10 &  0.58 &  \gradient{1.04} &  0.44 &  1.22 &  \gradient{1.09} &  1.61 &  2.22 &  \gradient{1.14} \\
RLV     &  0.10 &  0.58 &  \gradient{1.04} &  0.37 &  1.22 &  \gradient{1.07} &  1.36 &  2.21 &  \gradient{1.12} \\
RLG     &  0.12 &  0.60 &  \gradient{1.05} &  0.37 &  1.27 &  \gradient{1.06} &  1.53 &  2.36 &  \gradient{1.11} \\
NBI     &  0.14 &  0.84 &  \gradient{1.04} &  0.46 &  1.73 &  \gradient{1.05} &  0.61 &  3.14 &  \gradient{0.98} \\
\hline
\multicolumn{10}{l}{\textbf{S\&P 500}} \\
\hline
S5COND  &  0.14 &  0.67 &  \gradient{1.06} &  0.29 &  1.46 &  \gradient{1.03} &  0.82 &  2.68 &  \gradient{1.01} \\
S5CONS  &  0.07 &  0.44 &  \gradient{1.04} &  0.41 &  0.95 &  \gradient{1.10} &  1.09 &  1.79 &  \gradient{1.09} \\
S5ENRS  &  0.09 &  0.88 &  \gradient{1.02} &  0.51 &  1.83 &  \gradient{1.06} &  0.98 &  3.36 &  \gradient{1.02} \\
S5FINL  &  0.07 &  0.81 &  \gradient{1.02} &  0.43 &  1.70 &  \gradient{1.07} &  1.55 &  2.93 &  \gradient{1.12} \\
S5HLTH  &  0.10 &  0.54 &  \gradient{1.04} &  0.44 &  1.17 &  \gradient{1.09} &  1.07 &  2.08 &  \gradient{1.06} \\
S5INFT  &  0.15 &  0.74 &  \gradient{1.05} &  0.57 &  1.58 &  \gradient{1.08} &  1.97 &  2.99 &  \gradient{1.11} \\
S5MATR  &  0.14 &  0.75 &  \gradient{1.05} &  0.58 &  1.56 &  \gradient{1.10} &  0.94 &  2.93 &  \gradient{1.03} \\
S5RLST  &  0.14 &  0.80 &  \gradient{1.05} &  0.45 &  1.64 &  \gradient{1.07} &  1.21 &  2.87 &  \gradient{1.07} \\
S5TELS  &  0.09 &  0.67 &  \gradient{1.04} & -0.11 &  1.44 &  \gradient{0.97} &  0.85 &  2.79 &  \gradient{1.08} \\
S5UTIL  &  0.20 &  0.58 &  \gradient{1.11} &  0.30 &  1.26 &  \gradient{1.05} &  1.56 &  2.17 &  \gradient{1.15} \\
S5INDU  &  0.13 &  0.65 &  \gradient{1.05} &  0.34 &  1.42 &  \gradient{1.05} &  1.09 &  2.59 &  \gradient{1.06} \\
\hline
\multicolumn{10}{l}{\textbf{Europe}} \\
\hline
DAX     &  0.15 &  0.70 &  \gradient{1.06} &  0.63 &  1.53 &  \gradient{1.11} &  1.60 &  2.85 &  \gradient{1.11} \\
CAC     &  0.10 &  0.68 &  \gradient{1.04} &  0.57 &  1.46 &  \gradient{1.12} &  1.28 &  2.61 &  \gradient{1.12} \\
BEL20   &  0.08 &  0.61 &  \gradient{1.04} &  0.68 &  1.33 &  \gradient{1.17} &  1.69 &  2.35 &  \gradient{1.19} \\
UKX     &  0.09 &  0.56 &  \gradient{1.05} &  0.42 &  1.15 &  \gradient{1.11} &  1.14 &  1.99 &  \gradient{1.14} \\
IBEX    &  0.12 &  0.70 &  \gradient{1.06} &  0.70 &  1.55 &  \gradient{1.15} &  1.16 &  2.74 &  \gradient{1.11} \\
KFX     &  0.15 &  0.64 &  \gradient{1.06} &  0.93 &  1.33 &  \gradient{1.20} &  1.28 &  2.58 &  \gradient{1.05} \\
OMX     &  0.11 &  0.68 &  \gradient{1.04} &  0.56 &  1.41 &  \gradient{1.11} &  0.80 &  2.55 &  \gradient{1.04} \\
SMI     &  0.10 &  0.55 &  \gradient{1.06} &  0.62 &  1.14 &  \gradient{1.17} &  1.39 &  1.94 &  \gradient{1.18} \\
\hline
\multicolumn{10}{l}{\textbf{APAC}} \\
\hline
AS51    &  0.123 &  0.50 &  \gradient{1.08} &  0.49 &  1.04 &  \gradient{1.14} &  1.54 &  1.96 &  \gradient{1.18} \\
\hline
\multicolumn{10}{l}{\textbf{Japan}} \\
\hline
NKY     &  0.09 &  0.72 &  \gradient{1.03} &  0.61 &  1.53 &  \gradient{1.11} &  1.07 &  2.73 &  \gradient{1.06} \\
TPX     &  0.09 &  0.66 &  \gradient{1.04} &  0.61 &  1.41 &  \gradient{1.13} &  1.00 &  2.52 &  \gradient{1.07} \\
\hline
\multicolumn{10}{l}{\textbf{BRIC}} \\
\hline
IBOV    &  0.13 &  0.88 &  \gradient{1.03} &  0.64 &  1.89 &  \gradient{1.07} &  0.53 &  3.46 &  \gradient{0.96} \\
RTSI    &  0.24 &  1.00 &  \gradient{1.07} &  0.75 &  2.32 &  \gradient{1.08} &  1.10 &  4.36 &  \gradient{1.01} \\
NIFTY   &  0.15 &  0.67 &  \gradient{1.05} &  0.80 &  1.47 &  \gradient{1.12} &  0.93 &  3.14 &  \gradient{0.97} \\
MXIN    &  0.18 &  0.66 &  \gradient{1.06} &  0.87 &  1.47 &  \gradient{1.14} &  0.91 &  3.09 &  \gradient{0.97} \\
SHCOMP  &  0.08 &  0.73 &  \gradient{1.03} &  0.15 &  1.67 &  \gradient{1.01} &  0.31 &  3.48 &  \gradient{0.98} \\
SHSZ300 &  0.11 &  0.81 &  \gradient{1.03} &  0.40 &  1.85 &  \gradient{1.04} &  0.57 &  3.77 &  \gradient{0.98} \\
\hline
\end{tabular}
\end{adjustbox}
\caption{Location $\mu$, scale $\sigma$, and asymmetry $\kappa$ parameters of ALD for daily, weekly and monthly returns (in percent) obtained from the MCMC fits. Asymmetry values are highlighted with gradient colors varying from min (yellow) to max (red) values.}
\label{table:ALD-fit}
\end{table}

\section{Univariate portfolio allocation rules}
\label{sec:allocation-ald}

In Section ~\ref{sec:empirical}, we showed that the ALD, which has both fat tails and skewness, provides a good fit for the empirical data. Given the distribution $P(x)$ and the utility function $U(x,a)$, we can now determine the optimal portfolio weights. 

\subsection{Univariate Gaussian returns}
\label{subsec:univ-gauss}

To develop intuition, we first discuss the mean-variance framework, which assumes Gaussian returns. Given that the portfolio return $x$ follows a normal distribution $x \sim N(\mu,\sigma)$, the expected utility function $E_N[U_a(x,a)]$ can be expressed as:
\be
E_N[U_a(x,a)]=\int d x\,\, U_a(x,a) N(\mu,\sigma)=\frac{1-e^{\frac{1}{2} a^2 \sigma ^2- a \mu }}{a}
\label{eq:expectedutility}
\ee
where we examine the CARA utility function $U_a(x,a)$ from Eq.~\ref{eq:utility}. According to Eq.~\ref{eq:potfolio-opt}, the optimal portfolio maximizes the expected utility function in Eq.~\ref{eq:expectedutility}. This is equivalent to maximizing the expression $\mu-\frac{1}{2} a \sigma ^2$.

Suppose there is one dollar to invest for a period of $t$ days. There are two possible investment choices: a risky asset with a random return $r_t$ that follows a probability distribution $N(\mu,\sigma)$, and a riskless asset with a fixed return $r_0$. Let the portfolio allocation weight assigned to the risky asset be $w$ and the weight assigned to the riskless asset be $(1-w)$.
Then, the investment result $x$ can be expressed as: 
\be
x=w(1+r_t)+(1-w)(1+r_0)=1 + r_0 + w(r_t - r_0)
\label{x-form}
\ee
This expression follows a normal distribution with mean $\mu_x = 1+r_0+w(\mu-r_0)$ and standard deviation $\sigma_x=w\sigma$, denoted by $N(\mu_x,\sigma_x)$. Maximizing $\mu_x - \frac{1}{2} a \sigma_x^2$ leads to the optimization problem:
\be
w^*_N= \arg \max_w[w(\mu-r_0)-\frac{a}{2} w^2 \sigma^2]
\label{eq:mark-opt-w}
\ee
and the optimal weight $w_N$ is given by the Markowitz formula:
\be
w^*_N=\frac{\mu-r_0}{a \sigma^2}
\label{mark-w}
\ee
Note that weight $w^*_N$ grows linearly for large $\mu$ without saturation.

\subsection{Univariate ALD returns}
\label{subsec:univ-ald}
To compute the expected utility function with ALD, we substitute its PDF into Eq.~\ref{eq:potfolio-opt}. The resulting explicit formula is cumbersome, but the optimization-relevant part can be extracted using the following formal analogy:
\begin{equation}
\begin{aligned}
w^* = \arg \max_w E_P[U_a(x,a)] \equiv
 \left.  \arg \max_w  \big( - E_P[e^{i t x}]\right\vert_{t=i a}\big)
\end{aligned}
\end{equation}
The last term is 
a characteristic function (CF) $E_P[e^{i t x}]$ of the probability distribution $P$, well known for most statistical distributions. For the ALD the CF is given by~\cite{Kotz:2001}:
\be
\phi(t;\mu,\sigma,\mu_a) = \frac{\exp \left(i \mu t\right)}{1 +\frac{\sigma^2 t}{2}-i \mu_a t}
, \,\,\,\mu_a=\frac{\sigma}{\sqrt{2}} (\frac{1}{\kappa} -\kappa)
\ee
and results in the following mean $E[X]=\mu+\mu_a$ and variance $Var[X]=\sigma^2+\mu_a^2$. For any random variable $X\sim ALD(\mu,\sigma,\kappa)$ the following property holds $c+X\sim ALD(c+\mu,\sigma,\kappa)$. Thus, assuming the outcome of investment is given by Eq.~\ref{x-form}, the portfolio return is distributed according to $ALD(1 + r_0 + w(\mu - r_0), w \sigma,\kappa)$. The optimal weight $w^*_{ALD}$ is given by:
\be
w^*_{ALD}=\arg \max_w  \frac{-e^{-a w (\mu-r_0)}}{1-\frac{a^2 w^2 \sigma^2 }{2}+a w \mu_a},\,\,\,\,\, s.t. \,\, \frac{a^2 w^2 \sigma^2}{2}-a w \mu_a<1
\label{ald-weight}
\ee
Taking the logarithm, Eq. \ref{ald-weight} can be rewritten as follows:
\be
w^*_{ALD}=\arg \max_w \left[ w (\mu-r_0)+\frac{1}{a} \log\left(1-\frac{a^2}{2} w^2 \sigma^2+a w \mu_a\right)\right] ,\, s.t. \,\, \frac{a^2 w^2 \sigma^2}{2}-a w \mu_a<1
\label{eq:ald-optweight}
\ee
In the symmetrical case $\kappa=1$, the constraint condition simplifies to $|w|<\frac{\sqrt 2}{\sigma a}$. If $a \rightarrow 0$, we arrive to the mean-variance objective function by Taylor expansion. 

In order to avoid trivial cluttering, we assume later in the text that an investor chooses between stock and cash ($r_0$=0). 
Minimization of the objective function in Eq.~\ref{eq:ald-optweight} leads to the quadratic equation on weights: 
 \be
 a^2 \kappa  \mu  \sigma ^2 w^2+a \sigma  \left(\sqrt{2} \left(\kappa ^2-1\right) \mu  +2 \kappa  \sigma \right)  w+\sqrt{2} \left(\kappa ^2-1\right) \sigma -2
   \kappa  \mu=0
 \ee
which has the following (positive) solution:
\be
w^*_{ALD}=\frac{\sqrt{2 \left(\kappa ^2+1\right)^2 \mu ^2+4 \kappa ^2 \sigma ^2}-\sqrt{2} \left(\kappa ^2-1\right) \mu -2 \kappa  \sigma }{2 a \kappa  \mu  \sigma }
\label{ald-weight-full}
\ee
In the limit of $\frac{\mu}{\sigma} \to 0$, Eq.~\ref{ald-weight-full} simplifies to
\be
w^*_{ALD}=\frac{1}{a \sigma}\left(-\frac{\kappa ^2-1}{\sqrt{2} \kappa  }+\frac{\left(\kappa ^2+1\right)^2}{4 \kappa ^2 }\frac{\mu}{\sigma}\right)+O\left(\frac{\mu}{\sigma}\right)^2
\label{eq-ald-2}
\ee
Additionally, if $\kappa  \to 1$, we get
\be
w^*_{ALD}=-\sqrt 2 \frac{(\kappa -1)}{a \sigma}+\frac{\mu}{ a \sigma^2}+O\left(\frac{\mu}{\sigma}\right)^2
\ee
The second item is the standard mean-variance weight from Eq.~\ref{mark-w}. In the presence of long only constraint $w > 0$, the solution only allows allocation to the risky asset if $\mu>\sqrt 2 (\kappa -1) \sigma$. For SPX, the monthly skewness $\kappa_M=1.13$, and we obtain $\mu_M>\sqrt 2 (\kappa_M -1) \sigma_M=0.18*\sigma_M$.

Interestingly, the first term in Eq.~\ref{eq-ald-2} associated with the skewness of the distribution appears to be small, but not negligible, relative to the second term, which resembles the Markowitz solution from Eq.~\ref{mark-w}. Taking parameters for S\&P 500 daily returns from Table~\ref{table:ALD-fit}, we get
\be
-\frac{\kappa ^2-1}{\sqrt{2} \kappa }=-0.06, \,\,\,  \frac{\left(\kappa ^2+1\right)^2}{4  \kappa ^2 }\frac{\mu}{\sigma}=0.17
\ee  
We see that even a very small skewness $\kappa=1.042$ adds a contribution  comparable to the mean-variance contribution. 

The large $\frac{\mu}{\sigma}$ expansion leads to the following asymptotic solution for optimal portfolio weights:
\be  
w^*_{ALD}=\frac{\sqrt 2}{a \sigma \kappa }+O\left(\frac{1}{\mu}\right) 
\label{eq:univ-ald-large-mu}
\ee
As can be seen, the weights in Eq.~\ref{eq:univ-ald-large-mu} do not grow indefinitely, as in the Markowitz solution~\ref{mark-w}, but saturate and are independent of the strength of the signal. This allocation rule can be relevant for quantitative strategies with a high signal-to-noise ratio.

\subsection{Marginalization over location parameter \texorpdfstring{$\mu$}{}}
\label{subsec:margin-mu}

The location parameter $\mu$ is often not predictable. To model it, we assume that $\mu$ follows a normal distribution $\mu \sim N(\mu_0,\sigma_0)$ with parameters $\mu_0$ and $\sigma_0$, and marginalize (integrate) Eq.~\ref{eq:potfolio-opt} over location parameter $\mu$. The optimal weight $w^*$ is given by:
\be
w^*=\arg \max_w \int dx\,\, \int d \mu \,\, U(x,a) P(x;\mu,\sigma) N(\mu;\mu_0,\sigma_0)
\label{formula-port}
\ee
where $P(x)$ is the distribution of portfolio returns $x$.

To derive the optimal weight for both Gaussian and ALD returns, we rely on the following identity: $\int_{-\infty}^\infty d\mu\, e^{-a w \mu}  \, N(\mu;\mu_0,\sigma_0)=e^{\frac{1}{2} a^2 \sigma_0^2 w^2-a \mu_0 w}$.

For normally distributed returns, the optimal weight $w^*_N$ after marginalization over $\mu$ is given by:
\be
w^*_{N} = \arg \max_w\left[ (-1) e^{-a (\mu_0-r) w+\frac{a^2}{2} w^2 (\sigma^2+\sigma_0^2) }\right]
\label{norm-w-margin}
\ee 
Taking logarithm of Eq. \ref{norm-w-margin}, we arrive to:  
\be 
w^*_N = \arg \max_w \left[w(\mu_0 - r_0)-\frac{a}{2} w^2 (\sigma^2+\sigma_0^2)\right] = \frac{\mu_0-r_0}{a (\sigma^2+\sigma_0^2)}
\label{mark-w-margin}
\ee
For ALD returns, the optimal weight $w^*$ is:
\be
w^*_{ALD} = \arg \max_w  \frac{-e^{\frac{1}{2} a^2 \sigma_0^2 w^2-a (\mu_0-r) w}}{1-\frac{a^2 w^2 \sigma^2 }{2}+a w \mu_a} ,\,\,\,\,\, s.t. \,\, \frac{a^2 w^2 \sigma^2}{2}-a w \mu_a<1
\label{ald-w-margin}
\ee 
The optimal weight is given by a solution of cubic equation with respect to $w$:
\be
w^3 a^3 \kappa  \sigma ^2 \sigma_0^2 +
w^2 a^2  \left(\sqrt{2} \left(\kappa ^2-1\right) \sigma  \sigma_0^2-\kappa  \mu_0 \sigma^2\right)-
\ee
$$
-w a \left(\sqrt{2} \left(\kappa ^2-1\right) \mu_0 \sigma +2  \kappa(\sigma ^2+\sigma_0^2)\right)-\sqrt{2} \left(\kappa^2-1\right) \sigma +2 \kappa  \mu_0 =0
\label{ald-w-equation}
$$
The roots of the cubic equation can be found using Cardano's formula. Of the three solutions, only one has a negative second derivative with respect to $w$, which guarantees a maximum required in Eq.~\ref{ald-w-margin}. Furthermore, we do not provide a cumbersome formula for the exact solution, but study the three most important asymptotics. We assume that $\kappa \to 1$ in all cases. 

For investors with low confidence in the predicted expected return $\frac{\mu_0}{\sigma_0} \to 0$, the optimal weight is given by:
\be 
\lim_{\frac{\mu_0}{\sigma_0} \to 0} w^*=\frac{mean}{a (\sigma ^2+\sigma_0^2)}+O(\kappa-1)^2
,\,\, mean=\mu_0-\sqrt{2} (\kappa-1)\sigma 
\label{lowc}
\ee

For investors with high confidence in the predicted expected return $\frac{\mu_0}{\sigma_0} \to \infty$, the optimal weight is given by:
\be 
\lim_{\frac{\mu_0}{\sigma_0} \to \infty} w^*=\frac{\sqrt{2}}{a \sigma \kappa}+O(\kappa-1)^2\approx\frac{\sqrt{2}}{a \sigma}
\label{highc}
\ee

The limit of $\sigma_0 \to 0$ is given by:
\be 
\lim_{\sigma_0 \to 0} w^*=-\frac{\sqrt{2} (\kappa -1)}{a\sigma}+\frac{\sqrt{2\mu_0^2+\sigma^2}-\sigma}{a \mu_0 \sigma}+O(\kappa-1)^2
\label{lowsigma0}
\ee
 
\section{Multivariate case}
In the multivariate case of $N$ assets, the vectors $\boldsymbol  w,  \boldsymbol  \mu ,\boldsymbol  \mu_a \in R^N$ and covariance matrix $\boldsymbol  \Sigma \in R^{N \times N}$. 
In the case of multivariate Gaussian returns $\boldsymbol r_t \sim N(\boldsymbol \mu,\boldsymbol  \Sigma)$, the expected utility function is proportional to:
\be
E_N[U_a]\sim (-1) \exp(\frac{a^2}{2} \boldsymbol w^T \boldsymbol \Sigma \boldsymbol w -a \boldsymbol  \mu^T \boldsymbol  w)
\label{opt-markovitz1}
\ee
After taking the logarithm of $E_N[U_a]$, the optimization problem becomes:
\be
 \boldsymbol  w^*=\arg \max_{ \boldsymbol w} \left[ \boldsymbol  \mu^T \boldsymbol  w-\frac{a}{2} \boldsymbol w^T \boldsymbol \Sigma \boldsymbol w \right]
\label{opt-markovitz2}
\ee
which has the following solution: 
\be 
 \boldsymbol  w^*=\frac{1}{a} \boldsymbol  \Sigma^{-1} \boldsymbol  \mu 
\ee
Furthermore, we find allocation rules for portfolio returns distributed according to the multivariate ALD. 
The distribution is defined by its location vector $\boldsymbol{\mu}$, scale matrix $\mathbf{\Sigma}$, and asymmetry parameter $\boldsymbol{\mu_a}$.  The CF of multivariate ALD is given by \cite{Kotz:2001}: 
\be 
\phi(\boldsymbol  t;\boldsymbol  \mu, \boldsymbol \Sigma,\boldsymbol  \mu_a) = \frac{\exp \left(i \boldsymbol \mu^T \boldsymbol t\right)}{1 +\frac{1}{2} \boldsymbol  t^T \boldsymbol \Sigma \boldsymbol t-i \boldsymbol \mu_a^T \boldsymbol t}
\ee
Here,  $\boldsymbol \mu_a$ is still a scale-dependent quantity. There are multiple ways to make it scale-independent. We define $\boldsymbol \mu_a$  using scale-independent parameters $\boldsymbol \kappa$ as follows:
\be
\boldsymbol  \mu_a=\frac{1}{\sqrt 2}\sqrt{ Diag(\boldsymbol \Sigma)} (\boldsymbol \kappa^{-1}-\boldsymbol \kappa)
\ee
The mean and variance are given by: 
\be
Mean= \boldsymbol  \mu+\boldsymbol  \mu_a, \,\, Var=\boldsymbol  \Sigma+\boldsymbol  \mu_a^T \boldsymbol  \mu_a
\label{eq:ald-mean}
\ee
The expected utility function $E_{ALD}[U_a]$ is proportional to:
\be 
E_{ALD}[U_a] \sim \frac{-\exp \left(-a  \boldsymbol \mu^T \boldsymbol w\right)}{1 -\frac{a^2}{2} \boldsymbol  w^T \boldsymbol \Sigma \boldsymbol w+a \boldsymbol \mu_a^T \boldsymbol w}
\label{eq:33}
\ee
Taking the logarithm of  $E_{ALD}[U_a]$ and omitting irrelevant constants, we have the following optimization problem:
\be
 \boldsymbol w^*=\arg \max_{\boldsymbol w} \left[  \boldsymbol \mu^T \boldsymbol w+\frac{1}{a} \log\left(1-\frac{a^2}{2} \boldsymbol  w^T \boldsymbol \Sigma \boldsymbol w+a \boldsymbol \mu_a^T \boldsymbol w\right) \right]
\label{eq:multi-ald-weight}
\ee
such that $\frac{a^2}{2} \boldsymbol  w^T \boldsymbol \Sigma \boldsymbol w-a \boldsymbol \mu_a^T \boldsymbol w<1$.
Maximization of Eq.~\ref{eq:multi-ald-weight} results in the following system of quadratic equations with respect to $\boldsymbol w$: 
\be
\boldsymbol \mu  \left (1 -\frac{a^2}{2} \boldsymbol  w^T \boldsymbol \Sigma \boldsymbol w+a \boldsymbol \mu_a^T \boldsymbol w \right)+(\boldsymbol \mu_a-a \boldsymbol \Sigma \boldsymbol w)=0 
\label{eq:ald-q2}
\ee

\begin{figure}[h]
    \centering
    \includegraphics[width=0.6\textwidth]{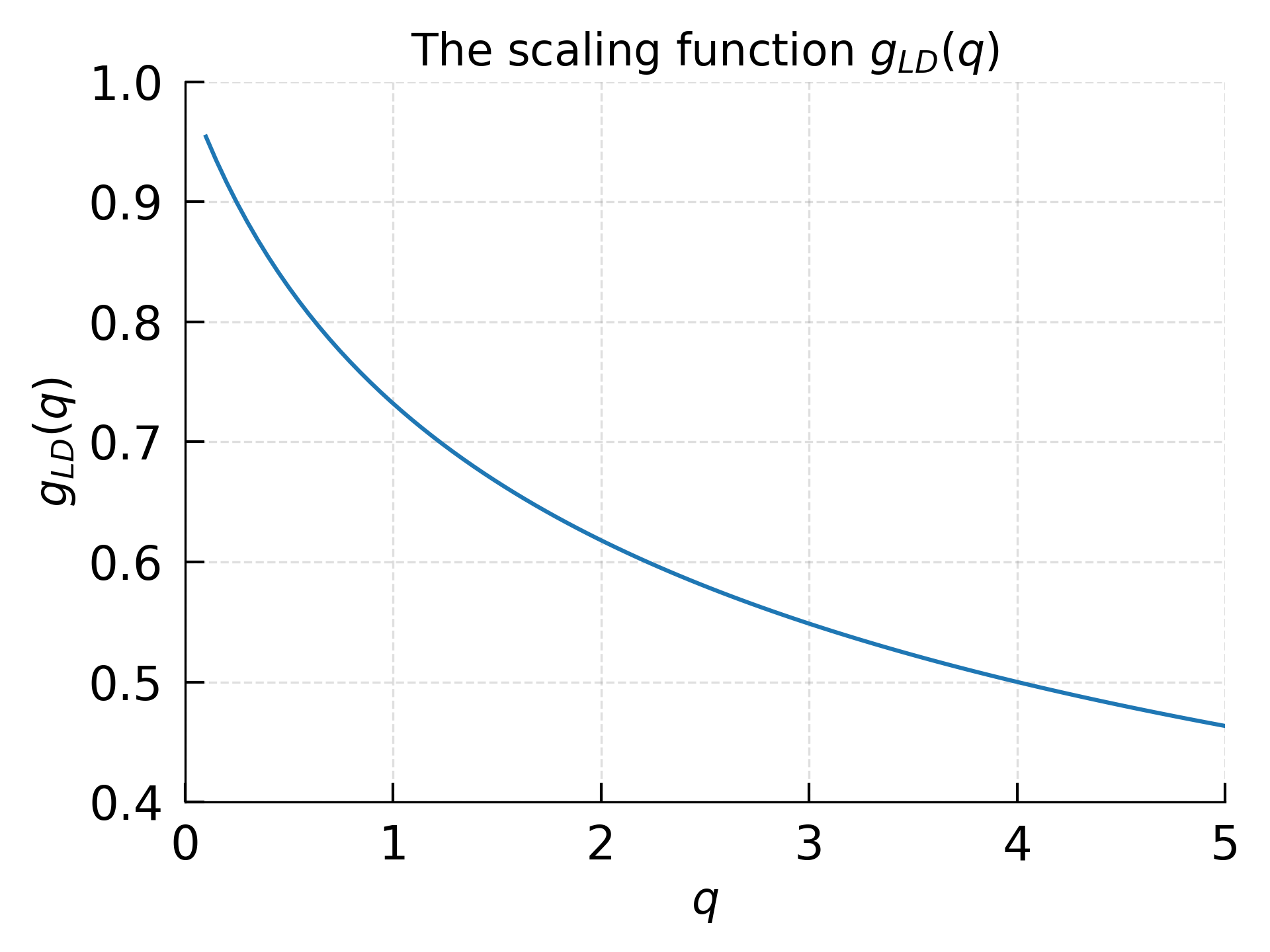}
\caption{The scaling function $g_{LD}(q,\alpha)$}
    \label{fig:my_label}
\end{figure}

For the symmetric case with $\boldsymbol \mu_a = 0$, the solution is given by:
\begin{equation}
\boldsymbol w^* = \frac{1 - \frac{a^2}{2} d}{a} \boldsymbol \Sigma^{-1} \boldsymbol \mu=g_{LD}(q,\alpha) \times \left(\frac{1}{a}\boldsymbol \Sigma^{-1} \boldsymbol \mu \right)
\label{eq:ald-q3}
\end{equation}
where $d = \boldsymbol w^T \boldsymbol \Sigma \boldsymbol w$ and $g_{LD}(q,\alpha)=1 - \frac{a^2}{2} d$. Substituting the solution \eqref{eq:ald-q3} into the definition of $d$, we obtain a quadratic equation with the following solution:
\begin{equation}
d = \frac{2 (1 + q - \sqrt{1 + 2 q})}{a^2 q}
\end{equation}
where $q = \boldsymbol \mu^T \boldsymbol \Sigma^{-1} \boldsymbol \mu$. We present the function $g_{LD}(q)$ in Figure~\ref{fig:my_label} for different values of $q$. The scaling function $g_{LD}(\alpha,a)$ is positive and decreases as the signal-to-noise level $q$ increases. 

For the one-dimensional case, the solution in Eq.\ref{eq:ald-q3} coincides with the solution in Eq.\ref{lowsigma0} when  $\kappa=1$.
In the general case, Eq.~\ref{eq:ald-q2} can be solved numerically. The \texttt{solve\char`_poly\char`_system} solver in the Sympy Python package is a reasonable choice for this task.

Similar to the univariate case considered in Section~\ref{subsec:margin-mu}, we study the expected utility function marginalized over $\boldsymbol \mu$ with $\boldsymbol\mu\sim N (\boldsymbol \mu_0,\boldsymbol \Sigma_0)$. Here, $\boldsymbol \Sigma_0$ is a diagonal matrix whose elements are equal to the variance of individual components of $\boldsymbol \mu$. To derive the optimal weight for both multivariate Gaussian and ALD returns, we rely on the following identity: $\int_{-\infty}^\infty d \boldsymbol \mu\, e^{-a \boldsymbol  w^T \boldsymbol  \mu}  \, N(\boldsymbol \mu; \boldsymbol \mu_0,\boldsymbol \Sigma_0)=e^{\frac{1}{2} a^2 \boldsymbol w^T \boldsymbol \Sigma_0 \boldsymbol w-a \boldsymbol \mu_0^T \boldsymbol w}$,

For normally distributed returns, the optimal weights $\boldsymbol  w^*$ after marginalization over $ \boldsymbol  \mu$ are given by:
\be
\boldsymbol w^*=\arg \max_{\boldsymbol w} \left[  -e^{\frac{1}{2} a^2 \boldsymbol w^T \boldsymbol (\boldsymbol  \Sigma+\boldsymbol  \Sigma_0) \boldsymbol w-a \boldsymbol \mu_0^T \boldsymbol w} \right]
\label{normal-weight-marg}
\ee
This equation can be rewritten as follows:
\be
\boldsymbol w^*=\arg \max_{\boldsymbol w} \left[  \boldsymbol \mu_0^T \boldsymbol w-\frac{a}{2}  \boldsymbol w^T \boldsymbol (\boldsymbol  \Sigma+\boldsymbol  \Sigma_0) \boldsymbol w \right]
\label{multi-normal-weight-r}
\ee
The optimal weights are given by:
\be
\boldsymbol w^*=\frac{1}{a}(\boldsymbol \Sigma+\boldsymbol \Sigma_0)^{-1}\boldsymbol \mu_0 
\label{eqsolution1norm}
\ee
The optimal weights $\boldsymbol w^*$ for ALD distributed returns are given by:
\be
\boldsymbol w^*=\arg \max_{\boldsymbol w}  \frac{-e^{\frac{1}{2} a^2 \boldsymbol w^T \boldsymbol \Sigma_0 \boldsymbol w-a \boldsymbol \mu_0^T \boldsymbol w}}{1-\frac{a^2 \boldsymbol w^T \boldsymbol \Sigma \boldsymbol w }{2}+a  \boldsymbol \mu_a^T \boldsymbol w}, \,\, s.t. \,\, \frac{a^2 \boldsymbol w^T \boldsymbol \Sigma \boldsymbol w }{2}-a \boldsymbol \mu_a^T \boldsymbol w <1
\label{ald-weight-marg}
\ee
This equation can be rewritten as follows:
\be
\boldsymbol w^*=\arg \max_{\boldsymbol w} \left[  \boldsymbol \mu_0^T \boldsymbol w-\frac{a}{2}  \boldsymbol w^T \boldsymbol \Sigma_0 \boldsymbol w+\frac{1}{a} \log\left(1-\frac{a^2}{2} \boldsymbol  w^T \boldsymbol \Sigma \boldsymbol w+a  \boldsymbol \mu_a^T \boldsymbol w\right) \right]
\label{multi-ald-weight-r}
\ee
such that $\frac{a^2}{2} \boldsymbol  w^T \boldsymbol \Sigma \boldsymbol w-a \boldsymbol \mu_a^T \boldsymbol w<1$.
Eq.~\ref{multi-ald-weight-r}  can be solved by a convex numerical optimizer such as CVXPY~\cite{Diamond:2016} even with additional constraints such as long-only, maximum number of position, turnover, and other constraints. 

Unconstrained solution of Eq.~\ref{multi-ald-weight-r} can be obtained by taking a derivative with respect of $\boldsymbol w$. Thus, we arrive at the following system of cubic equations:
\be
\boldsymbol (\boldsymbol \mu_0-a \boldsymbol \Sigma_0 \boldsymbol w)  \left (1 -\frac{a^2}{2} \boldsymbol  w^T \boldsymbol \Sigma \boldsymbol w+a \boldsymbol \mu_a^T \boldsymbol w \right)+(\boldsymbol \mu_a-a \boldsymbol \Sigma \boldsymbol w)=0 
\label{eqopt-mald}
\ee
For the symmetric case with $\boldsymbol  \mu_a = 0$, the solution is given by:
\begin{equation}
\boldsymbol w^* = \frac{(1-\frac{a^2}{2} d) \boldsymbol \mu_0}{a(  \boldsymbol \Sigma+(1-\frac{a^2}{2} d) \boldsymbol \Sigma_0)}
\label{eq:ald-marg-q3}
\end{equation}
where $d = \boldsymbol w^T \boldsymbol \Sigma \boldsymbol w$. Substituting the solution \eqref{eq:ald-marg-q3} into the definition of $d$, we obtain a nonlinear equation for d which can be solved numerically. Note that the role of uncertainty $\boldsymbol \Sigma_0$ of the location parameter $\boldsymbol \mu$ is similar to the regularization (shrinkage) approach \cite{Ledoit:2003}, but has a clearer interpretation. Moreover, there is no need to estimate a shrinkage constant.  In the general case, Eq.~\ref{eqopt-mald} can be solved numerically.

\section{Worst-case optimization}
One of the most prominent trends in trading is the increased use of leverage, both in stocks and in derivatives such as futures and options. Leverage enables traders to control a large contract value with a relatively small amount of capital. Leveraged trading often carries a non-negligible probability of ruin, as it amplifies both potential gains and losses. Consequently, to manage such risks, asset allocation under the worst-case outcome scenario often emerges as the optimal strategy in trading. In worst-case optimization, we maximize the expected value of the utility function under the worst-case outcome for the expected returns:
\begin{equation}
\boldsymbol w^*=\arg \max_{\boldsymbol w} \min_{\boldsymbol \mu \in D} E[U]
\end{equation}
where $D$ is the worst outcome under a model $\boldsymbol \mu \sim N(\boldsymbol \mu_0,\boldsymbol \Sigma_0)$.

\subsection{Worst-case optimization for the risk-neutral case}
If the risk component is neglected, the optimization problem is given by:

\begin{equation}
\boldsymbol w^*=\arg \max_{\boldsymbol w} \min_D \boldsymbol \mu^T \boldsymbol w,\,\, s.t. \sum_{i=1}^N w_i=1, \,w_i\ge 0
\label{maxmin-worst}
\end{equation}
We assume that the expected returns follow a multivariate normal distribution with mean $\boldsymbol \mu_0$ and diagonal variance $\boldsymbol \Sigma_0$. Given that $\boldsymbol \mu \sim N(\boldsymbol \mu_0,\boldsymbol \Sigma_0) = \boldsymbol \mu_0+\boldsymbol Y$ and $\boldsymbol Y\sim N(0,\boldsymbol \Sigma_0)$, the worst outcome for the weighted sum is given by:
\begin{equation}
\min_D \boldsymbol \mu^T \boldsymbol w = \boldsymbol \mu_0^T \boldsymbol w + \min \boldsymbol Y *\max_i w_i
\label{min-worst-rn}
\end{equation}
where $\max_i w$ is the largest weight from $\boldsymbol w$.
We estimate the minimal value of error $\min \boldsymbol Y$ under the assumption that $\boldsymbol Y\sim N(0,\boldsymbol \Sigma_0)$. Because matrix $\boldsymbol \Sigma_0$ is diagonal, all $Y_i$ are independent, and the problem is to find the expected value  $\langle Y_{min} \rangle$ of the minimum of N random variables $Y_i\sim N(0,\sigma^0_i)$ with Gaussian CDF $F_i(Y_i\le t)$. The CDF of the minimum of such variables is given by:

\begin{equation}
F_Y(y) = P(\min_i Y_i \le t) = 1 - \prod_{i}^N(1 - F_i(Y_i\le t))
\end{equation}
The PDF and the expected value of the minimum are given by:
\begin{equation}
f_Y(y) = \frac{dF_Y(y)}{dy} = \sum_{i=1}^N f_i(y) * \prod_{i \neq i} (1 - F_j(y)),\,\, \langle Y_{min} \rangle = \int_{-\infty}^{-\infty} dy \, y f_Y(y)
\label{worst-y-pdf}
\end{equation}
the expected value $\langle Y_{min} \rangle$ can be evaluated numerically by substituting into Eq.~\ref{worst-y-pdf} the probability density function $f(x)$ and cumulative distribution function $F(x)$ of a normal distribution with mean $\mu$ and variance $\sigma^2$:
\begin{equation}
f(x) = \frac{1}{\sigma\sqrt{2\pi}} e^{-\frac{(x - \mu)^2}{2\sigma^2}},\,\,\,
F(x) = \frac{1}{2} \left[ 1 + \operatorname{erf} \left( \frac{x - \mu}{\sigma\sqrt{2}} \right) \right]
\end{equation}
where $\operatorname{erf}(x)$ is the error function. In the case of equal $\sigma^0_i=\tilde \sigma_0$ and large N, the $\langle Y_{min}\rangle$ has the form $\langle Y_{min}\rangle \approx -\tilde \sigma_0 \sqrt{2 \log N}$.

Eq.~\ref{min-worst-rn} is a convex problem, and any solution satisfying the Karush-Kuhn-Tucker (KKT) conditions can be expected to be optimal. The KKT method generalizes the Lagrange multiplier method in the presence of inequality. When the objective function and constraints are convex, the KKT conditions are not only necessary but also sufficient for global optimality.

As we show in Appendix A, the solution of Eq.~\ref{min-worst-rn}, derived with Lagrange multipliers and KKT conditions, exhibits a simple behavior. For small $\langle Y_{min} \rangle$, all weights are allocated to an asset with the largest $\mu^0_i$. As $\langle Y_{min} \rangle$ starts to increase, the allocations change to a uniform allocation for $k$ assets with the largest $\mu^0_i$, where $k$ is an implicit function of  $\boldsymbol \mu^0$ and $\boldsymbol \sigma^0$. For large $\langle Y_{min} \rangle$, under the condition $|\langle Y_{min}\rangle|/N> \left( \langle \boldsymbol \mu^0 \rangle-\min_i \mu^0_i\right)$, the solution is an \textit{equal weights} $\frac{1}{N}$ allocation. This condition is satisfied early if the difference between the average $\langle \boldsymbol \mu^0 \rangle$ and the minimal expected return $\min_i \mu^0_i$ is small, which typically occurs for similar assets. The solution can be understood qualitatively, as the loss function in Eq.~\ref{min-worst-rn} is dominated by the $\max_i w$ term, which penalizes any allocation deviating from a uniform $1/N$. The $\max_i w$ term can also be interpreted as an infinity norm of the vector $\boldsymbol w$, given the following identity: $\max_i w = |w|_{\infty}$.

\subsection{Worst-case optimization with risk for the Gaussian return model}
If returns are distributed according to a multivariate normal distribution, we optimize Eq. \ref{opt-markovitz2} with the worst expected returns from Eq. \ref{min-worst-rn}:
\begin{equation}
\boldsymbol w^*=\arg \max_w\left[ \boldsymbol \mu_0^T \boldsymbol w+\min \boldsymbol  Y *\max_i w_i -\frac{a}{2} \boldsymbol w^T \Sigma \boldsymbol w\right], \,\, s.t. \sum_{i=1}^N w_i=1, \, w_i\ge 0
\label{gauss-min-worst-optimal}
\end{equation}
The solution can be obtained KKT conditions. For most practical cases, numerical optimization with CVXPY can provide a globally optimal solution with minimal effort.

Eq. \ref{gauss-min-worst-optimal} has an important limiting case when the contribution $\boldsymbol \mu_0^T \boldsymbol w$ can be neglected. In this case, we have:
\begin{equation}
\boldsymbol w^*=\arg \min_w \left[\max_i w_i + \frac{b}{2} \boldsymbol w^T \Sigma \boldsymbol w\right], \,\, s.t. \sum_{1}^N w_i=1, \,w_i\ge 0,\,\, b>0
\label{mm-min-worst}
\end{equation}
where $b=\frac{a}{|\min \boldsymbol  Y|}$. Because both the terms are convex, the optimization problem is convex, which ensures the existence of a unique global optimum. We discuss KKT solution of Eq. \ref{mm-min-worst} in Appendix B.

It is instructive to analyze two terms in the objective function Eq. ~\ref{mm-min-worst} separately. The first term, $\max_i w_i$, penalizes the maximum weight in the portfolio. Minimizing this term alone would lead to equal-weighted or more evenly distributed portfolio allocations. The second term, $\boldsymbol w^T \Sigma \boldsymbol w$, can be thought of as a measure of portfolio variance. Minimizing this term alone would lead to a minimum variance portfolio. When both terms are combined, the optimization problem aims to balance the trade-off between minimizing portfolio variance and avoiding concentrated portfolio allocations. As the constant term $b$ decreases, the optimization problem will put more emphasis on avoiding concentrated allocations, leading to a more evenly distributed portfolio.

A portfolio with similar characteristics is a risk parity(RP) portfolio, which is also known to provide a trade-off between the minimum variance and equally weighted portfolios \cite{Maillard:2010}. In the case of equal correlation, RP allocation is equal volatility allocation. The RP objective function is not convex, and to compute the optimal weights, nonlinear optimization methods, such as sequential quadratic programming, are required. Most importantly, the RP approach misses the critical ingredient: any information about expected returns and their associated uncertainties.

Our formulation offers a convex alternative to RP and includes an adjustable parameter $b$ that governs the trade-off between portfolio volatility risk and the worst-case scenario for the expected return prediction. The information about expected return mean  $\boldsymbol \mu_0$ can also be included. We refer to this class of portfolios as minimax (MM) portfolios.

We propose a heuristic rule to find a starting value for parameter $b$ that balances the lack of excessive concentration and correlation structure of a portfolio. Let $S$ be the entropy of portfolio weights $S=-\sum_{i=1}^N w_i \log w_i$ and  $N_{eff}$ is the effective number of securities in a portfolio $N_{eff}=\exp{S}$. A plausible $b^*$ is given by the condition that the entropy of the MM portfolio is equal to the average of entropy for the equal weight portfolio $S_{eq}$ and the minimum variance portfolio $S_{mv}$:
\begin{equation}
S_{b^*}=\frac{1}{2}(S_{eq}+S_{mv})
\end{equation}
This rule provides a starting point for the parameter $b$, which can be further refined through numerical simulations or other optimization techniques. Alternatively,  an investor can choose $b^*$ to obtain a specific value of $N_{eff}(b^*)/N$.  $N_{eff}=N$ for an equally weighted portfolio and  $N_{eff}=1$ for a single stock portfolio. 

\subsection{Worst case optimization with risk for the ALD return model}
Assuming that returns are distributed according to a multivariate ALD, we optimize Eq.~\ref{eq:multi-ald-weight} with the worst expected returns from Eq.~\ref{min-worst-rn}:
\be
\boldsymbol w^* =\arg \max_{\boldsymbol w} \left[\boldsymbol \mu_0^T \boldsymbol w+\min Y *\max_i w_i+\frac{1}{a} \log\left(1-\frac{a^2}{2}  \boldsymbol  w^T  \boldsymbol \Sigma \boldsymbol  w+a \boldsymbol\mu_a^T \boldsymbol  w \right) \right]
\label{ald-min-worst}
\ee
$$
 s.t. \sum_{i=1}^N w_i=1, \,w_i\ge 0, \,\, \,\, \frac{a^2}{2}  \boldsymbol  w^T  \boldsymbol \Sigma \boldsymbol  w-a \boldsymbol \mu_a^T \boldsymbol  w <1
$$
For general combinations of parameters, numerical optimization with CVXPY can provide a global optimum for Eq.~\ref{ald-min-worst} in minimal time.

If the contribution of expected return $\boldsymbol  \mu_0^T \boldsymbol  w$ is small, the optimization function can be rewritten as follows: 
\be
\boldsymbol w^*  =\arg \min_w \left[ \max_i w_i-b \log\left(1-\frac{a^2}{2}  \boldsymbol  w^T  \boldsymbol \Sigma \boldsymbol  w+ a \boldsymbol  \mu_a^T \boldsymbol  w \right) \right] 
\label{wc-ald-weight}
\ee
$$
 s.t. \sum_{i=1}^N w_i=1, \,w_i\ge 0, \,\, \,\, \frac{a^2}{2}  \boldsymbol  w^T  \boldsymbol \Sigma \boldsymbol  w-a \boldsymbol  w^T \boldsymbol \mu_a<1
$$
Empirically, we observe that the solution of Eq.~\ref{wc-ald-weight} is often close to the one defined by the Gaussian case in Eq.~\ref{mm-min-worst}. 

\section{Imposing graphical structure on a covariance matrix}
\label{sec:blocks}
Covariance and correlation coefficients are frequently used to measure the relationship between stocks.
A high correlation coefficient may arise either because the stocks are directly correlated with one another or because they are correlated with a similar index through their beta values. In the latter case, the index acts as a confounder, and when controlling for index returns, the stocks may move independently. Mathematically, this means that the stocks are conditionally independent, given the return of the index. Covariance is subject to numerical noise arising from finite samples and the existence of confounding variables. Moreover, spurious correlations are abundant in financial time series \cite{Leinweber:2007}.

Although machine learning algorithms, such as graphical lasso \cite{Friedman:2008}, can be used to learn the true conditional dependence (graphical) structure, experienced portfolio managers likely have this structure in mind before allocating real capital to an investment strategy. As a first approximation, the information on conditional independence can be incorporated into portfolio construction by utilizing a block structure derived from shared characteristics, such as sector, investment style, geographical region, or asset class. This structure assumes a stronger relationship within blocks and weaker or no causal relationship between some blocks, simplifying portfolio risk estimation and allocation. As a result, many spurious correlations between blocks disappear. Our objective is to estimate the covariance and precision matrices with a superimposed structure (adjacency) matrix $\boldsymbol A$.

In this section, we examine how to impose a graphical structure on covariance or precision matrices, which often results in a block structure. Although a component of the covariance matrix $\boldsymbol \Sigma$ represents \textit{marginal dependence} between corresponding random variables, elements of the precision matrix $\boldsymbol \Theta$ signify their \textit{conditional independence}, providing information about partial correlations between each pair of variables after accounting for the effects of all other variables. The elements of the partial correlation matrix $\varrho_{ij}$ can be obtained from the precision matrix $\Theta$ as follows: $\varrho_{ij}=-\frac{\Theta_{ij}}{\sqrt{\Theta_{ii} \Theta_{jj}}}$. The off-diagonal elements of $\varrho$ can be used as an adjacency matrix of the portfolio graphical structure.

Consider a portfolio consisting of two energy sector stocks, $E_1$ and $E_2$, and two healthcare stocks, $H_1$ and $H_2$. The returns of stocks in different sectors are conditionally independent; if we are interested in the return of $E_1$ and know the returns of $E_2$, the information on healthcare stocks returns becomes redundant. In other words, the block diagonal structure implies that there is no linear relationship between the assets when controlling for the other assets in the portfolio. 

 We demonstrate the concept using the Graphical Sparse Precision Matrix Estimation (GSPME) method, as implemented in the GraphSPME package~\cite{Lunde:2022}. This approach combines the precision matrix estimation routine with respect to a graph \cite{Le:2021} with the automatic and adaptive shrinkage of sample covariance matrices developed in \cite{Touloumis:2015}. Importantly, the resulting precision matrix estimates are guaranteed to be positive semi-definite, which allows for the utilization of efficient matrix factorization methods, such as Cholesky decomposition, in order to perform matrix inversion and subsequently obtain a covariance matrix.

As an example, we consider a portfolio consisting of Exxon Mobil (XOM) and Chevron (CVX) energy stocks, and Pfizer (PFE) and Merck (MRK) healthcare stocks, using daily returns for the year 2022. Assuming conditional independence between the energy and healthcare sectors, the structure information about conditional independence is encoded by an adjacency matrix $A$, which contains zeros and ones:
$$
\boldsymbol A=\left(
\begin{tabular}{rrrr}
        1 &        1 &             0 &           0 \\
        1 &        1 &             0 &           0 \\
        0 &        0 &             1 &           1 \\
        0 &        0 &             1 &           1 \\
\end{tabular}
\right)
$$
We compare covariance $\boldsymbol \Sigma$  and precision matrices $\boldsymbol \Theta$  and correlation $\boldsymbol Corr$ and partial correlation $\boldsymbol \varrho$ matrices using direct calculation and the proposed method for the portfolio above in Appendix C. 

The problem of inverting a covariance matrix becomes severe when the number of assets $N$ is larger than the number of observations $d$ as the matrix becomes singular. The graphical structure reduces the number of nonzero elements in the covariance matrix, thereby improving the condition number and making the matrix inversion
operation more stable.
In summary, we believe that the precision matrix $\boldsymbol \Theta$ is a \textit{primary} object for problems that require block (top-down) constraints, and the covariance matrix $\boldsymbol \Sigma$ can be derived from $\boldsymbol \Theta$ through inversion operations.

Another source of noise in covariance matrix estimation is the non-stationarity of return time series. The noise encompasses two components: variance and correlation structure. In order to mitigate these effects, it is prudent to employ a separation strategy as proposed in \cite{Barnard:2000}, where standard deviations and correlations are modeled independently. The covariance is decomposed as follows:
\begin{equation}
\boldsymbol \Sigma=\boldsymbol \Lambda \boldsymbol R\boldsymbol \Lambda
\end{equation}
where, $\boldsymbol \Lambda$ represents a diagonal matrix containing volatility elements $\sigma_i$ and $R$ denotes the corresponding correlation matrix. For volatility estimation in $\boldsymbol \Lambda$, one can utilize forward-looking volatility estimations derived from option markets. The correlation matrix $\boldsymbol R$ remains relatively stable, except during regime changes such as market crashes when all assets become highly correlated and move in unison.

\section{Portfolio allocation for long-term investors}
The long-term \mbox{\it{cross-sectional}} returns of major stock indexes are well approximated by log-normal distribution~\cite{Markov:2023}. We speculate that perception of risk of long-term investors is better described by the power-law CRRA utility function. The CRRA utility function measures risk from proportional rather than absolute wealth changes, keeping risk aversion consistent at different wealth levels. Application of CRRA to short-term trading is limited due to volatility drag effect and thus CARA utility is preferred.

The expected utility function is given by substitution of the CRRA utility $U_r(x,\gamma)$ and log-normal distribution function $P(x) = LN(x; \mu,\sigma)$ into Eq.~\ref{eq:potfolio-opt}:
\begin{align}
\begin{split}
& E_{LN}[U_r(x,\gamma)]=\int d x\,\, U_r(x,\gamma) \, LN(x;\mu,\sigma)=\frac{e^{(1-\gamma)\mu+(1-\gamma)^2\frac{\sigma^2}{2}}-1}{1-\gamma} \\ & E_{LN}[U_r(x,\gamma=1)] = \mu
\end{split}
\label{eq-logn}
\end{align}
The maximization of $E_{LN}[U_r(x,\gamma)]$ has a clear interpretation for $\gamma=0$ and $\gamma=1$. In the former case, the utility function is linear (risk-neutral) $U_r(x,0)=x-1$ and an investor should maximize the average return of the portfolio:
\be
w^*_{LN} = \max e^{\mu+\frac{\sigma^2}{2}}=\max E[x]
\ee
In the latter case, the utility function is logarithmic $U_r(x,1)=\ln(x)$, and an investor should maximize the median return of the portfolio:
\be
w^*_{LN} = \max e^{\mu}=\max median[x]
\ee
The maximization of the wealth median which coincides with geometric average for log-normal distribution of outcome also appears in Kelly criterion and in "Just one more" paradox in finance. 

Assume that each stock price $X^i$ in a portfolio follows a geometric Brownian motion:
\be
\frac{dX^i_{t+1}}{X^i_t} = \mu_i\, dt + \sigma\,dW_t,
\label{eq:log-bm}
\ee
with the percentage drift distributed according to the normal distribution $\mu_i\sim N(\mu_d,\sigma_d)$. In this model ~\cite{Markov:2023}, the total return $\rho_i=X^i_T/X^i_0$ of a randomly chosen stock $i$ at time $T$ follows a log-normal distribution with $\mu_m=\mu_d T -\frac{1}{2}\sigma^2 T$ and  $\sigma_m^2=\sigma^2 T+\sigma_d^2 T^2$. 
The mean and median of the return distribution are given by:
\be
mean[\rho]=e^{\mu_m+\frac{\sigma_m^2}{2}} =e^{\mu_d T+\frac{1}{2} \sigma_d^2T^2}
\ee
\be
median[\rho]=e^{\mu_m}=e^{ \mu_d T-\frac{1}{2}\sigma^2 T}
\ee
A risk-neutral investor with utility function  $U_r(x,0)=x-1$ should maximize the expected value of the drift $\mu_d$ and, most importantly, maximize the dispersion of the drift $\sigma_d$ by taking diversified risky bets and disregarding the volatility of the stock in the portfolio. Drift dispersion ${\sigma_d}^2T^2$ represents the effect of  continuous compounding of winners pushing the average return up, while a large body of distribution is concentrated around the mode and goes to zero.  In turn, an investor with logarithmic utility $U_r(x,1)=\ln(x)$ should maximize the expected value of the drift $\mu_d$ while minimizing the volatility exposure $\sigma^2 T$. 

 The CRRA utility can be marginalized to account for location and scale uncertainty if the uncertainty of the parameters needs to be considered. Assuming the location parameter $\mu$ is modeled as a Gaussian random variable $\mu \sim N(\mu_0, \sigma_0^2)$ and the scale parameter $\sigma^2$  is distributed according to a gamma distribution $\sigma^2\sim \Gamma(\frac{\alpha}{2}, \frac{2 s^2}{\alpha})$, we have:
\be 
 E_{\mu,\sigma^2}[U_r]=\int_{0}^{\infty} d s^2 \int_{-\infty}^{\infty} d \mu \,\, E_{LN}[U_r(x,\gamma)] N(\mu;\mu_0,\sigma_0^2) \Gamma(\sigma^2;\frac{\alpha}{2}, \frac{2 s^2}{\alpha})
\ee
Using the following identities:
\be 
\int_{-\infty}^{\infty} \,\, d\mu \, E_{LN}[U_r(x,\gamma)] N(\mu;\mu_0,\sigma_0^2) =\frac{e^{\frac{g^2}{2} \left(\sigma ^2+\sigma_0^2\right)+g \mu_0}-1}{g}, \,\,
\ee
\be 
\int_{0}^{\infty} \,\, d\sigma^2 \, e^{\frac{1}{2} g^2 \sigma^2}\Gamma(\sigma^2;\frac{\alpha}{2}, \frac{2 s^2}{\alpha})=\frac{1}{\left(1-\frac{g^2}{\alpha} s^2\right)^\frac{\alpha}{2}} 
\ee
 where $g=1-\gamma$. Consequently, the optimal portfolio construction equates to the following optimization problem for $0 \leq \gamma < 1$:
 \be 
 \max \left[ \mu_0+\frac{g}{2} \sigma_0^2-\frac{\alpha}{2 g} \log \left(1-\frac{g^2 s^2}{\alpha} \right)\right]
 \ee
The parameters $\mu_0$, $\sigma_0$, $s^2$, and $\alpha$ are derived from the steady-state distribution of Eq.~\ref{eq:log-bm}, which in turn is governed by the selection of securities with a particular distribution of drift and volatility.

\section{Conclusions}
\label{sec:conclusions}

This paper introduces a series of models designed to enhance the quality of asset allocation rules. By analyzing historical stock index returns across various time scales, we find that the ALD serves as a suitable model for daily, weekly, and monthly returns. We derive allocation rules for ALD returns and systematically compare them to the mean-variance approach. We also introduce a model where the expected returns are represented as Gaussian random variables and study the average and worst-case scenarios, thus taking into account the uncertainty associated with the expected returns. We argue that deriving the sample covariance matrix from the precision matrix is more natural if the graphical structure, which encodes the conditional independence of assets, is known. This procedure enhances the numerical stability of the covariance matrix inversion by removing spurious correlations. Finally, we develop allocation rules for log-normal distributed cross-sectional returns, using a power-law utility function, thereby providing insight into optimal long-term portfolio allocations.


\subsection*{Appendix A: Worst-case optimization in risk-neutral case}
The objective of the worst-case optimization in the risk-neutral case is given by the following equation:
\be
\boldsymbol w^*=\arg \max_w \left[ \boldsymbol \mu_0^T \boldsymbol w   - c \cdot r \right],\, s.t.\,\,  
w_i \geq 0 ,\sum_{i=1}^N w_i = 1,\,\, w_i \leq r 
\ee
We can define the Lagrangian function as follows:
\be
L(\boldsymbol w, \lambda, \boldsymbol \eta, \boldsymbol \nu) = \boldsymbol \mu_0^T \boldsymbol w  - c \cdot r + \lambda \cdot (1 - \sum_{i=1}^N w_i) + \sum_{i=1}^N \eta_i \cdot w_i - \sum_{i=1}^N \nu_i \cdot (w_i - r)
\ee
\text{KKT conditions}:
\begin{enumerate}
  \item \text{Stationarity}: \( \nabla_{ \boldsymbol  w} L =  \boldsymbol  \mu_0  - \lambda \cdot \mathbf{1} +  \boldsymbol  \eta -  \boldsymbol \nu = 0, \,\,\,\, \partial_r L=-c+ \mathbf{1}^T  \boldsymbol \nu =0\)
  \item \text{Primal feasibility}:  \( w_i \geq 0 ,\sum_{i=1}^N w_i = 1 ,w_i \leq r\)
  \item \text{Dual feasibility}: \(\eta_i \geq 0, \nu_i \geq 0\)
  \item \text{Complementary slackness}: \(\eta_i \cdot w_i = 0 ,\,\, \nu_i \cdot (w_i - r) = 0\)
\end{enumerate}
Let us assume that $\mu_0^i$ are positive and ordered in descending order. For non-zero $w_i$, we have the following equation:
\begin{equation}
\eta_i = 0, \nu_i = \mu_0^i - \lambda , \nu_i \ge 0
\label{eq:nu-eq}
\end{equation}
Using the second stationarity condition, the optimal value of $\lambda^*$ is given by $\sum_{i=1}^N (\mu_0^i - \lambda^*)^+ = c$, where $(x)^+=\max(0,x)$. This leads to the following condition for $\lambda^*$: $\mu_0^{k+1}<\lambda^*<\mu_0^{k}$. The number of non-zero weights $k$ is defined as: $\sum_{i=1}^k (\mu_0^i - \mu_0^k) < c$ and $\sum_{i=1}^{k+1} (\mu_0^i - \mu_0^{k+1}) > c$. Thus, $\lambda^*=\frac{1}{k}\sum_{i=1}^k \mu_0^i-\frac{c}{k}$. When $\nu_i > 0$, we have $w^*_i = r$. For $\eta_i = 0$ and $\nu_i = 0$, we have $0 \le w_i \le r$, but this condition is not feasible as it requires the exact identity $\mu_0^i = \lambda^*$. 
For $\eta_i > 0$, we have $w^*_i = 0$ and $\nu_i=0$. Eq. \ref{eq:nu-eq} gives $k$ non-zeros weights which satisfy $\mu^i_0 \ge \lambda^*$. Therefore, we have non-zero uniform weights $w^i = r$ for $i \in [1, k]$. We then apply the normalization condition $\sum_{i=1}^N w_i = 1$, which yields a uniform allocation solution with $r_i^* = \frac{1}{k}$ for $i \in [1, k]$. It is evident that $k = N$ for a sufficiently large $c$ such that $\sum_{i=1}^N (\mu_0^i - \mu_0^N) < c$, and it results in the uniform allocation $\frac{1}{N}$.  

\subsection*{Appendix B: Worst-case optimization of MM portfolio}
The objective for the worst-case optimization of MM portfolio is given by the following equation:
\be
 \boldsymbol  w^*=\arg \min_{ \boldsymbol w}\left[r + \frac{b}{2} \cdot  \boldsymbol  w^T  \boldsymbol  \Sigma  \boldsymbol  w\right]
,\,\, s.t. \, w_i \geq 0, \sum_{i=1}^N w_i = 1,w_i \leq r
\ee
The Lagrangian function is given by:
\be
L(w, \lambda, \eta, \nu) = r + \frac{b}{2} \cdot  \boldsymbol  w^T  \boldsymbol  \Sigma  \boldsymbol  w + \lambda \cdot (1 - \sum_{i=1}^N w_i) - \sum_{i=1}^N \eta_i \cdot w_i + \sum_{i=1}^N \nu_i \cdot (w_i - r)
\ee
\text{KKT conditions}:
\begin{enumerate}
  \item \text{Stationarity}: \(\nabla_{\boldsymbol w} L = b   \boldsymbol \Sigma  \boldsymbol  w - \lambda \cdot \mathbf{1} -  \boldsymbol \eta +  \boldsymbol  \nu = 0\),\,\,\,\, \(\partial_r L=1- \mathbf{1} \nu =0\)
  \item \text{Primal feasibility}: \( w_i \geq 0 ,\sum_{i=1}^N w_i = 1, w_i \leq c\)
  \item \text{Dual feasibility}: \(\eta_i \geq 0 ,\nu_i \geq 0\)
  \item \text{Complementary slackness}: \(\eta_i \cdot w_i = 0 ,\nu_i \cdot (w_i - r) = 0\)
\end{enumerate}
From stationarity condition, we have: 
\be 
\boldsymbol w=\frac{\boldsymbol \Sigma^{-1}}{b}\left (\lambda \cdot \mathbf{1}+\boldsymbol \eta -\boldsymbol\nu \right)
\label{w-worst-diag-sigma} 
\ee 
Complementary slackness condition has three scenarios:\\
$$ 
1.\,\, w_i = 0: \quad \eta_{i} \geq  0 \text{ and } \nu_{i} =0 
$$
In this scenario,  the stationarity condition implies: $(\boldsymbol \Sigma^{-1}(\lambda \mathbf{1} +\boldsymbol \eta))_i = 0$. Let's denote the i-th row of the inverse of the matrix $\boldsymbol \Sigma$ as $\boldsymbol \Omega_i=(\boldsymbol\Sigma^{-1})_i$. Then the condition is $\boldsymbol \Omega_i (\lambda \mathbf{1} +\boldsymbol \eta)=0$.
$$
2. \,\, 0 < w_i < r: \quad \eta_{i} = 0 \text{ and } \nu_{i} = 0
$$
The stationarity condition simplifies to $ w_i=\frac{\lambda}{b}\left (\boldsymbol \Omega_i \cdot \mathbf{1}\right)$.
$$
3. \,\, w_i = r: \quad \eta_{i} = 0 \text{ and } \nu_{1i} \geq 0
$$
The stationarity condition implies that $\boldsymbol \Omega_i(\lambda \mathbf{1} -\boldsymbol \nu) =r b$

It is instructive to analyze the optimal allocation for a diagonal matrix \newline $\boldsymbol \Sigma=diag(\sigma_1^2,\sigma_2^2,
\ldots,\sigma_N^2)$. Assume the assets are ordered in the order of  increasing volatility $\sigma_i$: $\sigma_1 \le \sigma_2,\ldots,\le \sigma_N$. According to the slackness condition, we have three groups of indexes: $i0$ corresponds to $w_{i0}=0$, $i+$ corresponds to $0<w_{i+}<r$ and $ir$ corresponds to $w_{ir}=r$. Thus, we have the following condition for $\boldsymbol \nu$ for each group of indexes:
\be
\nu_{ir}=\lambda-r b \sigma_{ir}^2 \text{ and } \nu_{ir}\ge 0;\,\,  \nu_{i0}=0 ; \,\, \nu_{i+}=0
\ee
The number $k$ of weights $w_i=r$ is given by the maximum $k$ which still satisfies $\lambda^* \ge r b \sigma_{k}^2$. The second stationarity condition $\sum_{i=1}^N \nu_i=1$ is $\sum_{i=1}^N (\lambda -r b \sigma_i^2)^+=1$, where $(x)^+=\max (0,x)$. Thus, $\lambda^*=\frac{1}{k}(1+ b r^* \sum_{i=1}^k \sigma_i^2)$. The conditions for $\boldsymbol \eta$ are:
\be
\eta_{ir}=0; \,,\, \eta_{i0}=-\lambda \text{ and } \eta_{i0}\ge 0  ;\,,\, \eta_{i+}=0
\ee
For diagonal matrix $\boldsymbol \Sigma$,  $\lambda^*$ is positive, i0 set is empty, and $\eta_i=0$ for all $i$.
The normalization condition $\sum_{i=1}^N w_i=1$ implies $k r^*+\lambda^* \sum_{i=k+1}^N \frac{1}{b} \frac{1}{\sigma_i^2}=1$.  Solving the above equations, we obtain the optimal values of $\lambda^*$, $r^*$ and $k$. The optimal weights are given by Eq.~\ref{w-worst-diag-sigma}. There are $N-k$ $i+$ weights which are inversely proportional to the variance of assets normalized by $\lambda^*$ $w^*_{i+}=\frac{\lambda^*}{b} \frac{1}{\sigma_{i+}^2}$, and $k$ $ir$ weights $w^*_{ir}=r^*$. The KKT-based solution impeccably aligns with the numerical resolution computed using CVXPY.

For general $\boldsymbol \Sigma$, we have three group of weights: $w_{i0}=0$, $w_{ir}=r$ and $0<w_{i+}<r$   $\boldsymbol  w_{i+}=\frac{\lambda*}{b}\left (\boldsymbol \Omega_{i+} \cdot \mathbf{1}\right)$ where the weights $\boldsymbol  w$ are a function of $\boldsymbol \Sigma$ and $b$ and are obtained via the solution of the KTT equations or direct solution obtained with the CVXPY solver.

\section*{Appendix C: Block constraints on a sample portfolio}
We examine a portfolio comprising Exxon Mobil (XOM) and Chevron (CVX) energy stocks, as well as Pfizer (PFE) and Merck (MRK) healthcare stocks, utilizing daily returns for the year 2022. Assuming conditional independence between the energy and healthcare sectors, the structural information is encoded by an adjacency matrix $\boldsymbol A$, containing zeros and ones:
$$
\boldsymbol A=\left(
\begin{tabular}{rrrr}
        1 &        1 &             0 &           0 \\
        1 &        1 &             0 &           0 \\
        0 &        0 &             1 &           1 \\
        0 &        0 &             1 &           1 \\
\end{tabular}
\right)
$$
We estimate the precision matrix $\boldsymbol \Theta_{\text{sample}}$ by inverting the sample covariance matrix $\boldsymbol \Theta_{\text{sample}}=\boldsymbol \Sigma_{\text{sample}}^{-1}$ and the model estimation of the precision matrix $\boldsymbol \Theta_{\text{model}}$ as defined in section \ref{sec:blocks} with adjacency matrix $\boldsymbol A$ as follows:
$$
\boldsymbol \Theta_{sample}=
\left(\begin{tabular}{rrrr}
8974 &  -8400 &   198 &  -311 \\
-8400 &  10335 &  -684 &  -160 \\
 198 &   -684 &  5075 & -3614 \\
-311 &   -160 & -3614 &  9188 \\
\end{tabular}
\right)
,\,
\boldsymbol \Theta_{model}=
\left(\begin{tabular}{rrrr}
8438 &  -7827 &     0.0 &     0.0 \\
-7827 &  9585 &     0.0 &     0.0 \\
0 &      0 &  4898 & -3581 \\
0 &      0 & -3581 &  8962 \\
\end{tabular}
\right)
$$
The sample covariance  $\boldsymbol \Sigma_{sample}$ and model covariance $\boldsymbol \Sigma_{model}=\boldsymbol \Theta_{model}^{-1}$ matrices (multiplied by 10,000 for readability) are as follows:
$$
\boldsymbol \Sigma_{sample}=
\left(\begin{tabular}{rrrr}
4.9 &  4.0 &  0.7 &  0.5 \\
4.0 &  4.3 &  0.8 &  0.5 \\
0.7 &  0.8 &  2.9 &  1.2 \\
0.5 &  0.5 &  1.2 &  1.6 \\
\end{tabular}
\right)
,\,\,
\boldsymbol \Sigma_{model}=
\left(\begin{tabular}{rrrr}
4.9 &  4.0 &  0.0 &  0.0 \\
4.0 &  4.3 &  0.0 &  0.0 \\
0.0 &  0.0 &  2.9 &  1.2 \\
0.0 &  0.0 &  1.2 &  1.6 \\
\end{tabular}
\right)
$$
The sample correlation  $Corr_{sample}$ and model correlation $Corr_{model}$ matrices are as follows:
$$
Corr_{sample}=
\left(\begin{tabular}{rrrr}
1.00 &  0.88 &  0.19 &  0.19 \\
0.88 &  1.00 &  0.23 &  0.20 \\
0.19 &  0.23 &  1.00 &  0.55 \\
0.19 &  0.20 &  0.55 &  1.00 \\
\end{tabular}
\right)
,\,\,
Corr_{model}=
\left(\begin{tabular}{rrrr}
1.00 &  0.87 &  0.00 &  0.00 \\
0.87 &  1.00 &  0.00 &  0.00 \\
0.00 &  0.00 &  1.00 &  0.54 \\
0.00 &  0.00 &  0.54 &  1.00 \\
\end{tabular}
\right)
$$
The sample partial correlation  $\boldsymbol \varrho_{sample}$ and the model partial correlation $\boldsymbol \varrho_{model}$ matrices are as follows:
$$
\boldsymbol \varrho_{sample}=
\left(\begin{tabular}{rrrr}
1.00 &  0.87 & -0.03 &  0.03 \\
0.87 &  1.00 &  0.09 &  0.02 \\
-0.03 &  0.09 &  1.00 &  0.53 \\
0.03 &  0.02 &  0.53 &  1.00 \\
\end{tabular}
\right)
,\,
\boldsymbol \varrho_{model}=
\left(\begin{tabular}{rrrr}
1.00 &  0.88 &  0.00 &  0.00 \\
0.88 &  1.00 &  0.00 &  0.00 \\
0.00 &  0.00 &  1.00 &  0.55 \\
0.00 &  0.00 &  0.55 &  1.00 \\
\end{tabular}
\right)
$$
The off-diagonal elements of $\boldsymbol \varrho_{\text{sample}}$ are smaller compared to the off-diagonal elements of $Corr_{\text{sample}}$. This finding is consistent with our hypothesis that incorporating sector information results in conditionally independent blocks within the portfolio.
\newpage

\subsection*{Appendix D: Regression for the ALD parameters}

\begin{figure}[h]
\centering
  \includegraphics[width=\textwidth]{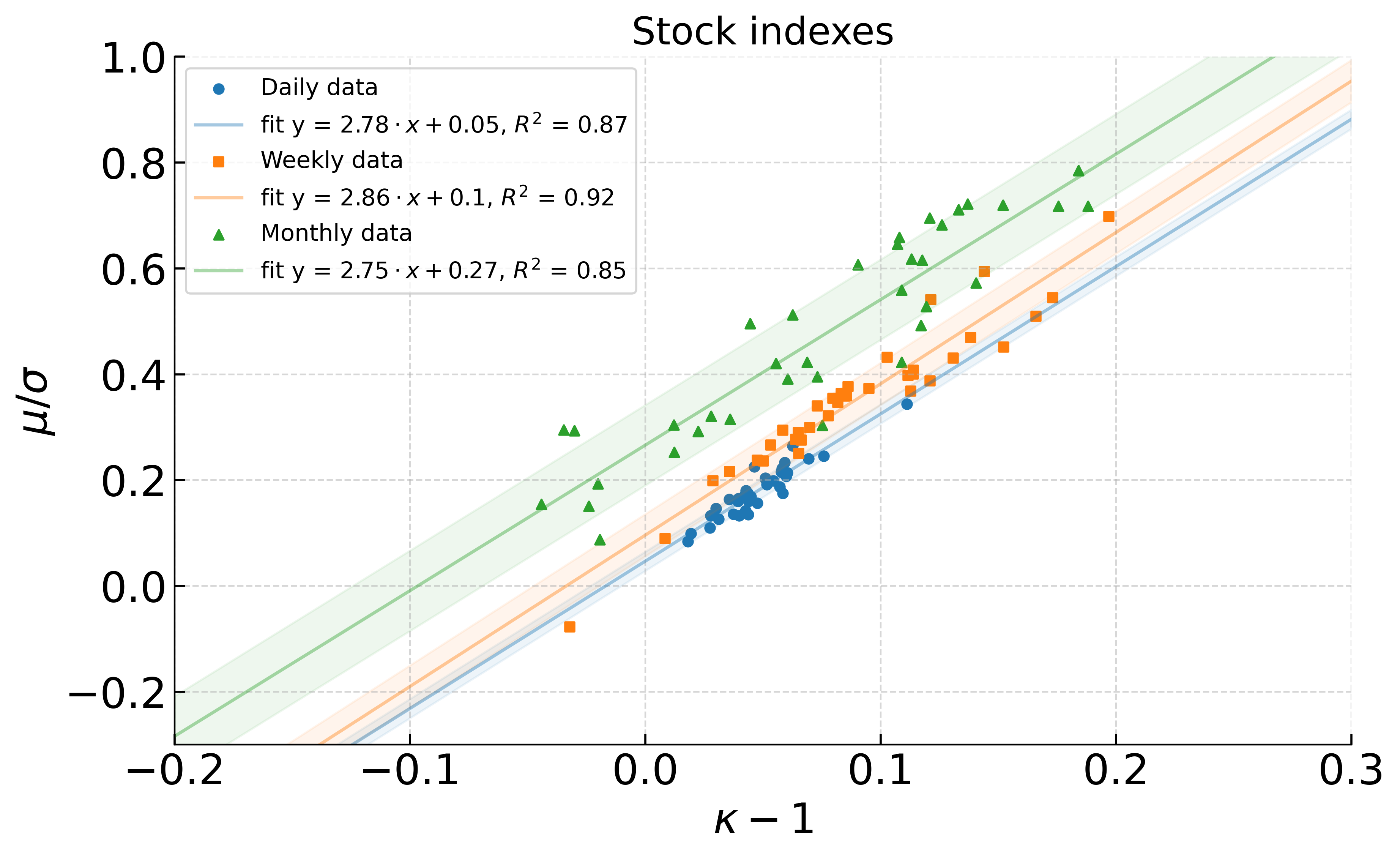}
  \label{fig:linear_fit_terms}
\caption{The empirical relation between the ALD parameters for daily (blue), weekly (orange), and monthly (green) returns is illustrated. The term $\frac{\mu}{\sigma}$ represents the location-to-scale ratio, while $\kappa-1$ represents the skewness of the distribution. Data points are fitted using a linear regression model, and the shaded region indicates the error bars. The quality of fit is assessed using the $R^2$ score, which is presented in the legend. It is noteworthy that the slope remains consistent across all timeframes, while the intercept demonstrates a monotonic increase.}
\end{figure}

\end{document}